\pdfoutput=1

\documentclass[12pt,a4paper,hyperref]{jhepLike}
\let\ifpdf\relax
\usepackage{ifpdf,color}
\let\normalcolor\relax
\usepackage{mathtools,shuffle,cite}
\newcommand{\eq}[1]{\vspace{-0.5pt}\begin{equation}#1\vspace{-0.5pt}\end{equation}}
\newcommand{\fwbox}[2]{\text{\makebox[#1][c]{$\hspace{-150pt}\displaystyle#2\hspace{-150pt}$}}}
\newcommand{\fwboxL}[2]{\text{\makebox[#1][l]{$#2$}}}
\newcommand{\fwboxR}[2]{\text{\makebox[#1][r]{$#2$}}}
\newcommand{\fig}[3]{\raisebox{#1}{\ \includegraphics[scale=#2]{#3}}}
\newcommand{\mi}{\raisebox{0.75pt}{\scalebox{0.75}{$\,-\,$}}}
\newcommand{\pl}{\raisebox{0.75pt}{\scalebox{0.75}{$\,+\,$}}}
\renewcommand{\phi}{\varphi}
\newcommand{\ab}[1]{\langle\hspace{-0.5pt}#1\hspace{-0.5pt}\rangle}
\newcommand{\bigger}[1]{\raisebox{-2.25pt}{\scalebox{1.75}{$#1$}}}
\newcommand{\Pfprime}{\text{Pf}\,'\!\hspace{1pt}}
\newcommand{\chy}{\Omega_{\mathrm{CHY}}}
\newcommand{\z}[2]{(\hspace{-0.5pt}#1,#2\hspace{-0.5pt})}
\newcommand{\zexpl}[2]{(z_{#1}\hspace{-0.5pt}\mi z_{#2}\hspace{-0.5pt})}
\newcommand{\sexpl}[2]{2\hspace{1pt}k_{#1}\hspace{-4.5pt}\cdot\hspace{-2.5pt}k_{#2}}
\newcommand{\ek}[2]{\epsilon\hspace{-0.5pt}k_{#1 #2}}
\newcommand{\ekexpl}[2]{2\hspace{1pt}\epsilon_{#1}\hspace{-4.5pt}\cdot\hspace{-2.5pt}k_{#2}}
\newcommand{\e}[1]{\epsilon_{#1}}
\newcommand{\eexpl}[2]{2\hspace{1pt}\epsilon_{#1}\hspace{-4.5pt}\cdot\hspace{-2.5pt}\epsilon_{#2}}

\title{{\LARGE \mbox{Manifesting Color-Kinematics Duality}}\\ {\LARGE\mbox{in the Scattering Equation Formalism}}}
\author{{\normalsize \mbox{N.~E.~J.~Bjerrum-Bohr$^1$, Jacob~L.~Bourjaily$^1$, Poul~H.~Damgaard$^1$, Bo~Feng\mbox{$^{1,2}$}}}\\
\mbox{{\mbox{$^1$}\ Niels Bohr International Academy and Discovery Center, University of Copenhagen}}\\
\mbox{{\mbox{$^{\phantom{1}}$}\ Blegdamsvej 17, DK-2100 Copenhagen \O, Denmark}}\\
\mbox{{\mbox{$^2$}\ Zhejiang Institute of Modern Physics, Zhejiang University}}\\
\mbox{{\mbox{$^{\phantom{2}}$}\ Hangzhou City, 310027, People's Republic of China}}\\\vspace{-5pt}
\mbox{\hspace{-13pt}{{\bf Email:} {\tt bjbohr@nbi.dk, bourjaily@nbi.ku.dk, phdamg@nbi.dk, fengbo@zju.edu.cn}}}
}
\date{\today}
\abstract{We prove that the scattering equation formalism for Yang-Mills amplitudes can be used to make manifest the theory's color-kinematics duality. This is achieved through a concrete reduction algorithm which renders this duality manifest term-by-term. The reduction follows from the recently derived set of identities for amplitudes expressed in the scattering equation formalism that are analogous to monodromy relations in string theory. A byproduct of our algorithm is a generalization of the identities among gravity and Yang-Mills amplitudes.}

\begin{document}
\vspace{-6pt}\section{Introduction}\label{sec:introduction}\vspace{-0pt}

The scattering equation formalism of Cachazo, He and Yuan (CHY) \cite{Cachazo:2013gna,Cachazo:2013hca,Cachazo:2013iea} has proven to provide a remarkably rich representation of amplitudes, giving us new and unusual tools with which to analyze quantum field theory. Based on a set of algebraic relations, the scattering equations, tree-level amplitudes for a variety of different quantum field theories are expressed in remarkably compact forms in any number of dimensions and for any number of external legs. Integrations over an auxiliary parameter space are required, but the integrals localize completely on the support of solutions to the scattering equations. Formally, this removes all integrations. For an $n$-point amplitude there are, however, $(n\mi3)!$ independent solutions and they must be summed over. This has been a bottleneck of the formalism, and much effort has gone into simplifying the unwieldy sum over solutions to the scattering equations \cite{Dolan:2014ega,Kalousios:2015fya,Cachazo:2015nwa,Baadsgaard:2015voa,Baadsgaard:2015ifa,Baadsgaard:2015hia,He:2015yua,Huang:2015yka,Sogaard:2015dba,Cardona:2015eba,Cardona:2015ouc,Dolan:2015iln,Gomez:2016bmv,Cardona:2016bpi,Zlotnikov:2016wtk}. A very simple solution to this problem was provided by the integration rules of \mbox{refs.\ \cite{Baadsgaard:2015voa,Baadsgaard:2015ifa,Baadsgaard:2015hia}}. Those rules provided a systematic way of evaluating the majority of integrals encountered in the CHY-formalism. Some additional integrals that appear in the case of Yang-Mills amplitudes had to be considered separately.

Recently, we have shown how to extend these integration rules to all cases needed for the evaluation of $n$-point Yang-Mills amplitudes \cite{Bjerrum-Bohr:2016juj}. The key to this development was to use the map between the CHY-formalism and string theory \cite{Bjerrum-Bohr:2014qwa} and apply the power of monodromy relations \cite{BjerrumBohr:2009rd,Stieberger:2009hq}. For the full amplitudes, these monodromy relations can neatly be arranged in two classes: by taking the real part of the amplitude relations one recovers, in the field theory limit, the KK-relations \cite{KK,DelDuca:1999rs} and a basis of amplitudes of size $(n\mi2)!$. Taking the imaginary part, one finds the BCJ-relations \cite{BCJ} and a further reduction to an $(n\mi3)!$ basis. In the CHY-formalism, the analog of these relations {\it on individual terms in the integrands} split up similarly \cite{Bjerrum-Bohr:2016juj}: the analog of the real parts of the relations yield useful algebraic identities, while the analog of the imaginary parts provide new non-trivial identities that are valid only on the support of the the scattering equations. This has been explored in further detail in \mbox{ref.\ \cite{Cardona:2016gon}.} 

The general solution for $n$-point Yang-Mills amplitudes given in \cite{Bjerrum-Bohr:2016juj} is provided in a form that does not make color-kinematics duality manifest. However, it is known indirectly that such a form must exist \cite{Cachazo:2013iea,Naculich:2014rta}. Now that we have the general integration rules available, we should thus be able to derive by direct computation the numerators that put the amplitude in BCJ-form, making the color-kinematics duality manifest. Indeed, as we shall show in detail below, the missing ingredient is precisely the new set of identities that follow from monodromy relations. 

As explained in \mbox{ref.\ \cite{Cachazo:2013iea}}, the color-kinematics duality of amplitudes in Yang-Mills theory will be made manifest so long as the (reduced) Pfaffian, $\Pfprime\Psi$, appearing in the CHY representation of Yang-Mills and gravity amplitudes can be expressed in terms that involve only single, permuted `Parke-Taylor' factors in the denominator---that is, terms with only a single Hamiltonian cycle. The corresponding integrals are then trivialized by means of the integration rules of \cite{Baadsgaard:2015voa,Baadsgaard:2015ifa,Baadsgaard:2015hia}, and in fact refer to trivalent Feynman graphs only \cite{Baadsgaard:2015ifa}. This is the form that we obtain using the reduction procedure described in this paper.

Our paper is organized as follows. In \mbox{section \ref{sec:review_of_scattering_equations}} we review the basic formalism and notation involved in the CHY representation of scattering amplitudes for Yang-Mills theory (and gravity). Amplitudes represented in this way involve a summation of terms arising from the (reduced) Pfaffian of a matrix denoted $\Psi$. Upon direct expansion, these terms are far from those needed to make the color-kinematics duality manifest. This fact and the form required to manifest duality is reviewed in \mbox{section \ref{subsec:colorkinematics}}. In \mbox{section \ref{sec:algorithm}} we describe how the monodromy relations can be used to systematically reduce terms of $\Pfprime\Psi$ so that the full Yang-Mills amplitude is provided explicitly in terms of a KK-basis with numerators satisfying Jacobi relations, as needed to achieve color-kinematics duality. We review the ingredients involved in \mbox{section \ref{subsec:operations}}, describe how these can be used to systematically reduce terms in \mbox{section \ref{subsec:algorithm}}, and prove that upon successive iterations of these reductions the result will always be of the form that makes color-kinematics duality manifest in \mbox{section \ref{subsec:algorithm_proof}}. In \mbox{section \ref{sec:examples}} we illustrate the results of this systematic algorithm in the case of four-particle scattering. And we conclude our work with a view towards applications of the same reduction algorithm to discover non-trivial identities for other theories.

\vspace{-0pt}\section{Review: Amplitudes in the Scattering Equation Formalism}\label{sec:review_of_scattering_equations}\vspace{-0pt}

Let us briefly review the essential ingredients for the representation of scattering amplitudes in the scattering equation formalism of CHY \cite{Cachazo:2013gna,Cachazo:2013hca,Cachazo:2013iea}. The essential observation is that on-shell momenta $k_a^{\mu}$ (in any number of dimensions) can be encoded in terms of auxiliary variables $z_a\!\in\!\mathbb{CP}^1$ via the {\it scattering equations}:
\eq{S_a\equiv\sum_{b\neq a}\frac{s_{ab}}{z_a-z_b}=0,\label{scattering_equations_definition}}
where $s_{ab}\!\equiv\!(k_a\pl k_b)^2\!=\!\sexpl{a}{b}$, and the index $a\!\in\!\{1,\ldots,n\}$ labels the external particles. Although there are $n$ equations, only $(n\mi3)$ are independent for momentum-conserving external momenta. This is reflected in an $\mathrm{SL}(2,\mathbb{C})$ invariance in the representation of momenta in terms of the auxiliary $z$ variables.

In terms of these auxiliary variables tree-level scattering amplitudes in any quantum field theory can be represented as integrals over the $z$'s, localized on the constraints (\ref{scattering_equations_definition}). Concretely, an $n$-particle scattering amplitude can always be represented in the form
\eq{\mathcal{A}_n=\int\!\chy\,\,\mathcal{I}(z)\,,\label{general_chy_amplitude}}
where $\mathcal{I}(z)$ depends on the theory in question, and $\chy$ is the ($\mathrm{SL}(2,\mathbb{C})$-fixed) volume-form on the space of the $z$'s, together with $\delta$-function constraints that impose the scattering equations:
\eq{\chy\equiv\frac{d^nz}{\mathrm{vol(SL(}2,\mathbb{C}))}\delta^{n-3}(S)\equiv\zexpl{i}{j}^2\zexpl{j}{k}^2\zexpl{k}{i}^2\!\!\prod_{a\notin\{i,j,k\}}\!\!dz_a\,\delta(S_a)\,.\label{chy_measure_definition}}
Because of the $\delta$-function constraints, the integral in (\ref{general_chy_amplitude}) is always fully localized on the solutions to the scattering equations, which number $(n\mi3)!$ in general. That is, integrals of the form (\ref{general_chy_amplitude}) can always be written simply as a sum:
\eq{\int\!\chy\,\,\mathcal{I}(z)=\sum_{\fwbox{40pt}{z^*|S_a(z^*)=0}}\mathcal{J}(z^*)\,\,\mathcal{I}(z^*)\,,\label{chy_integral_as_sum}}
where $\mathcal{J}$ is the Jacobian resulting from the $\delta$-function constraints. The details of this formula need not concern us here. But it is worth mentioning that finding ways to evaluate these integrals analytically {\it without} explicitly summing over the $(n\mi3)!$ solutions to the scattering equations has been the subject of much recent work.

For certain quantum field theories, the representation of amplitudes via (\ref{general_chy_amplitude}) takes an especially simple form (see, e.g.\ \cite{Cachazo:2014xea}). Perhaps the simplest is bi-adjoint scalar $\phi^3$-theory; in this case, amplitudes can be represented by
\eq{\mathcal{A}_n^{\phi^3}(1,2,\ldots,n)\equiv\int\!\chy\,\,\left(\frac{1}{\zexpl{1}{2}\zexpl{2}{3}\cdots\zexpl{n}{1}}\right)^2\,.\label{scalar_phi3_integrand}}
The factor being squared appearing above will play a recurring role in our work, and it is useful to give it a name. By analogy to the famous Parke-Taylor amplitude \cite{Parke:1986gb}, we define a cyclic sum of factors to be:
\eq{\hspace{5pt}\mathrm{PT}(1,\ldots,n)\equiv\frac{1}{\zexpl{1}{2}\zexpl{2}{3}\cdots\zexpl{n}{1}}\equiv\frac{1}{\z{1}{2}\z{2}{3}\cdots\z{n}{1}}\equiv\frac{1}{\ab{1\,2\cdots n}}.\hspace{-30pt}\label{parke_taylor_factor_defined}}
Here, we have also defined two bits of notation that will prove useful to us later:
\eq{\z{a}{b}\equiv\zexpl{a}{b}\quad\mathrm{and}\quad\ab{a\,b\cdots c}\equiv \left(\z{a}{b}\cdots\z{c}{a}\right)\,.\label{ab_and_zab_notation_defined}}

Color-ordered scattering amplitudes of Yang-Mills theory also have an especially simple CHY-representation:
\eq{\mathcal{A}_n^{\text{YM}}(1,2,\ldots,n)\equiv\int\!\chy\,\,\text{PT}(1,2,\ldots,n)\Pfprime\Psi(k,\epsilon)\,,\label{ordered_yang_mills_amplitude_in_chy}}
where $\Pfprime\Psi$ is the (reduced) Pfaffian of the matrix $\Psi$---the Pfaffian of the matrix $\Psi_{ij}^{ij}$ obtained by deleting two rows and columns $\{i,j\}$ (with $1 \leq i,j \leq n$) of $\Psi$:
\eq{\Pfprime\Psi\equiv(-1)^{i+j}\frac{1}{\zexpl{i}{j}}\text{Pf}(\Psi^{ij}_{ij})\quad\text{with}\quad\Psi\equiv\left(\hspace{-1pt}\begin{array}{c@{$\,\,\,\,\,\,\,$}c@{$\,\,\,$}}A&\,\fwboxR{0pt}{\text{--}\,}C^{\fwboxL{0pt}{T}}\\C&B\end{array}\hspace{0.5pt}\right)\,,\label{definition_of_reduced_pfaffian_and_of_psi}}
where the components of $\Psi$ are given by the matrices,
\eq{\begin{array}{llllll} A_{a\neq b}&\displaystyle\equiv\frac{s_{ab}}{\zexpl{a}{b}}\quad&B_{a\neq b}&\displaystyle\equiv\frac{\e{ab}}{\zexpl{a}{b}}\quad&C_{a\neq b}&\displaystyle\equiv\frac{\ek{a}{b}}{\zexpl{a}{b}}\\[14pt]
A_{aa}&\equiv0&B_{aa}&\equiv0&C_{aa}&\displaystyle\equiv \text{`}\frac{1}{\ab{a}}\text{'}\end{array}\label{components_of_psi_definition}}
where $s_{ab}\!\equiv\!\sexpl{a}{b}$, $\e{ab}\!\equiv\!\eexpl{a}{b}$, and $\ek{a}{b}\!\equiv\!\ekexpl{a}{b}$. The diagonal entries of the $C$-matrix---formally {defined} as `$1/\ab{a}$' above---will be made explicit in \mbox{section \ref{subsec:operations}} below (see equation (\ref{expansion_of_diagonal_entries_of_c_matrix_algebraic})); for now, we can simply consider it to be some abstract function of weight (-2) in $z_a$.

From the color-ordered partial amplitudes of Yang-Mills, it is easy to construct the full amplitudes. This can be done, for example, using a KK-representation \cite{KK,DelDuca:1999rs} involving a reduced basis of $(n\mi2)!$ color-ordered partial amplitudes:
\eq{\mathcal{A}_n^{\text{YM}}\equiv\sum_{\sigma\in\mathfrak{S}_{n-2}}c_{\{1,\sigma,n\}}\!\times\mathcal{A}_n^{\mathrm{YM}}(1,\sigma(2),\ldots,\sigma(n\mi1),n)\,,\label{full_yang_mills_from_ordered_amps}}
where the summation is over all permutations $\sigma$ of the set $\{2,\ldots,n\mi1\}$, and the color factors $c_{\{1,\sigma,n\}}$ are defined as contractions of the structure constants of the gauge group of the theory,
\eq{c_{\{1,\sigma,n\}}\equiv\sum_{\alpha_i}\left(f^{1\,\sigma(2)\,\alpha_1}\!f^{\alpha_1\,\sigma(3)\,\alpha_2}\!\cdots f^{\alpha_{n-3}\,\sigma(n-1)\,n}\right)\,.\label{definition_of_kk_color_factors}}

Finally, graviton scattering amplitudes can be represented in a way that is remarkably (and suggestively) similar to the
Yang-Mills squaring of KLT-relations \cite{Kawai:1985xq}:\\[-10pt]
\eq{\mathcal{A}_n^{\text{GR}}\equiv\int\!\chy\,\,(\Pfprime\Psi)^2\,.\label{gravity_amplitudes_in_chy}}
It may be worth mentioning that this remarkable representation of gravitational amplitudes, written this way first by Cachazo, He, and Yuan in \cite{Cachazo:2013gna}, is a natural generalization of the formula discovered by Hodges in \cite{Hodges:2012ym}.

\vspace{-0pt}\subsection{Color-Kinematics Duality in the CHY Representation}\label{subsec:colorkinematics}\vspace{-0pt}

A comparison between the representations of amplitudes in Yang-Mills (\ref{ordered_yang_mills_amplitude_in_chy}) and gravity (\ref{gravity_amplitudes_in_chy}), combined with the expansion into color factors in (\ref{full_yang_mills_from_ordered_amps}) is suggestively close to making the KLT relations between the two theories manifest. This immediately indicates that the CHY-formalism automatically incorporates color-kinematics duality. Indeed, as noticed by Cachazo, He and Yuan in \mbox{ref.\ \cite{Cachazo:2013iea}} (see also \cite{Monteiro:2013rya}), this correspondence can be made quite explicit. Using KLT-orthogonality \cite{BjerrumBohr:2010ta,BjerrumBohr:2010hn}, it follows that there exists an expansion of $\Pfprime\Psi$ of the form,
\eq{\Pfprime\Psi=\sum_{\sigma\in\mathfrak{S}_{n-2}}n_{\{1,\sigma,n\}}\!\times\text{PT}(1,\sigma(2),\ldots,\sigma(n\mi1),n) \,,\label{kk_expansion_of_pfaffian}}
where the kinematic factors $n_{\{1,\sigma,n\}}$ could only then be determined implicitly---as a sum over the color-ordered amplitudes, weighted by the KLT momentum kernel. As shown in \mbox{ref.\ \cite{Cachazo:2013iea}}, consistency between the Yang-Mills (\ref{ordered_yang_mills_amplitude_in_chy}) and gravity (\ref{gravity_amplitudes_in_chy}) can be achieved by the numerators $n_{\{1,\sigma,n\}}$ satisfying Jacobi relations. Color-kinematics duality in the CHY formalism can thus be made manifest once we have explicitly expanded the $\Pfprime\Psi$ according to (\ref{kk_expansion_of_pfaffian}), but a procedure for doing this has not been known until now. An explicit construction of the expansion (\ref{kk_expansion_of_pfaffian}) is what we provide in this note---providing a representation of Yang-Mills amplitudes in the CHY formalism  explicitly involving BCJ numerators, making color-kinematics duality manifest. 

It is perhaps not very well appreciated that Jacobi relations for numerator factors are not the most general solutions that follow from imposing the exact BCJ amplitude relations (which, in turn, follow from monodromy in string theory). There is more freedom in choosing numerators, completely consistent with all established amplitude relations \cite{Tye:2010dd,BjerrumBohr:2010zs}. In \mbox{ref.\ \cite{BjerrumBohr:2010zs}} general numerator identities were explored in detail.

\newpage
\vspace{-0pt}\section{\mbox{The Algorithm: Manifesting the Color-Kinematics Duality}}\label{sec:algorithm}\vspace{-0pt}

Every term in the expansion of the (reduced) Pfaffian $\Pfprime\Psi$ has manifest weight (-$2$) in all the $z$-variables---without any factors appearing in the numerator (when the diagonal terms from the $C$-matrix are left abstract). When represented as an oriented graph connecting nodes $a\!\to\!b$ for each factor $\z{a}{b}\!\equiv\!\zexpl{a}{b}$ appearing in the denominator, all nodes will necessarily lie along some closed (Hamiltonian) cycle. Thus, every term will correspond to a graph involving a collection of disconnected cycles---including one-cycles for each term arising from the diagonal of $C$.

We would like to systematically reduce every such term into ones with only a single cycle.  This can be achieved through the iterated use of simple cross-ratio and Schouten-like identities derived in \cite{Bjerrum-Bohr:2016juj} (see also \cite{Cardona:2016gon}). We will also make use of these these identities to expand the diagonal
terms from $C$ as suggested in \cite{Bjerrum-Bohr:2016juj}, in a way which directly parallels the other operations involved.

\vspace{-0pt}\subsection{Fundamental Operations, Identities, and Diagrammatic Notation}\label{subsec:operations}\vspace{-0pt}

The principal non-trivial identity that we will need for reduction is the monodromy  relation, which can be expressed in terms of cross ratios of factors $\z{a}{b}$ as follows. For any subset $A\!\subset\{1,\ldots,n\}$ with $2\!\leq\!|A|\!\leq\!(n\mi2)$ and any point $a\!\in\!A$, for each $b\!\in\!A^c$ we have the following identity \cite{Cardona:2016gon}:
\eq{1=-\!\sum_{\substack{\alpha\in A^{\phantom{c}}\\\beta\in A^c}}\!\frac{s_{\alpha\beta}}{s_A}\frac{\z{a}{\alpha}\z{\beta}{b}}{\z{b}{a}\z{\alpha}{\beta}}\,.\label{fundamental_cross_ratio_identity}}
(Here, $s_{A}\!\equiv\!(k_{a_1}\!\pl\cdots\pl k_{a_m})^2$ for any set $A\!\equiv\!\{a_1,\ldots,a_m\}$.) Strictly speaking, this `identity' is valid only on the support of the scattering equations. Nevertheless, because we are only interested in integrals supported on the scattering equations (using the form $\Omega_{\text{CHY}}$, which includes the $\delta$-functions imposing these constraints), we are free to consider equation (\ref{fundamental_cross_ratio_identity}) an actual identity for our purposes.

Our reduction procedure will only make use of (\ref{fundamental_cross_ratio_identity}) when the subset $A$ is a closed Hamiltonian cycle. Thus, we can represent this diagrammatically as follows:
\eq{\hspace{-0pt}1=-\!\sum_{\substack{\alpha\in A^{\phantom{c}}\\\beta\in A^c}}\!\frac{s_{\alpha\beta}}{s_A}\fwboxL{75pt}{\fig{-27.14pt}{1}{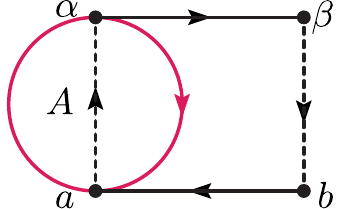}}\hspace{-0pt}\label{cross_ratio_identity_diagrammatic}}
Notice that we have used dashed lines to represent the factors of (\ref{fundamental_cross_ratio_identity}) appearing in the numerator---with arrows to indicate the signs of terms. Also, we have written the cycle $A$ in red to indicate that its size is arbitrary, and there may be many (implicit) points along it besides $a$ and $\alpha$. Lines denoting single factors $\z{a}{b}$ will always be drawn in black.

Any terms generated in this way can always be expanded into those with fewer numerators through the use of the so-called `KK' relations:
\eq{\text{PT}(a,A_1,\alpha,A_2)=(-1)^{|A_2|}\sum_{\substack{\\[-2pt]\fwbox{0pt}{\hspace{-0pt}\sigma\!\in\!(A_1\!\shuffle\!{A_2^R})}}}\;\;\text{PT}(a,\sigma,\alpha)\,,\label{kk_relations_algebraic}}
where for any set $A\!\equiv\!\{a_1,\ldots,a_m\}$, $\text{PT}(A)$ denotes the Parke-Taylor factor corresponding to the cycle $A$, $1/(\z{a_1}{a_2}\cdots\z{a_m}{a_1})$, ${A_2^R}$ denotes the subset $A_2$ with reversed ordering, and the summation is over all `shuffles' $\sigma$ of the sets $A_1$ and ${A_2^R}$. (Recall that these are simply permutations of $A_1\!\cup\!{A_2^R}$ which preserve the ordering of the two sets.) Diagrammatically, the KK-relations imply the following identity:
\eq{\hspace{-0pt}\fwboxR{0pt}{\fig{-27.14pt}{1}{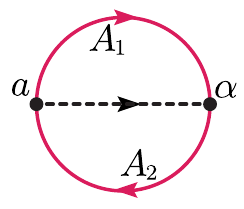}}=-(-1)^{|A_2|}\sum_{\substack{\\[-2pt]\fwbox{0pt}{\hspace{-0pt}\sigma\!\in\!(A_1\!\shuffle\!{A_2^R})}}}\fwboxL{0pt}{\fig{-27.14pt}{1}{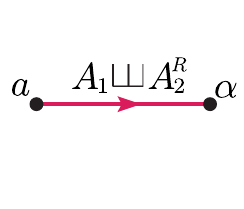}}\hspace{-0pt}\label{kk_relations_diagrammatic}}
Here, the meaning of the open chain should be obvious: it simply corresponds to the product of the factors $\z{a}{b}$ in the denominator for each directed edge from $a\!\to\!b$. Notice the extra minus sign appearing above---due to the relative sign between $\z{a}{\alpha}$ and the corresponding factor that would have arisen for the PT-factor.

Although the KK-relations may appear non-trivial, they are in fact strictly algebraic identities at the level of the integrand---independent of the support of the scattering equations, and valid for cycles $A$ of arbitrary length. Indeed, they can be derived through the iterated use of the simple (Schouten-like) identity:
\eq{\frac{\z{c}{a}}{\z{a}{b}\z{b}{c}}=\frac{\z{d}{a}}{\z{a}{b}\z{b}{d}}+\frac{\z{c}{d}}{\z{b}{c}\z{d}{b}},}
which can be represented diagrammatically as follows:
\eq{\hspace{10pt}\fwboxR{0pt}{\fig{-27.14pt}{1}{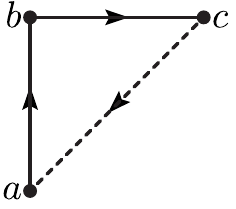}}=\fwboxL{70pt}{\fig{-27.14pt}{1}{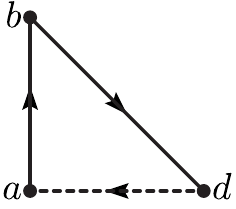}}+\hspace{-10pt}\fwboxL{00pt}{\fig{-27.14pt}{1}{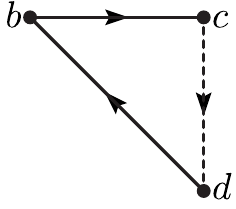}}\label{schouten_triangulation_figure}}
This identity is itself fairly trivial; but it is not hard to see how it can be iteratively used to obtain the familiar KK-relations stated above, and their application to (\ref{kk_relations_diagrammatic}).

It will be useful to always use the identity (\ref{kk_relations_diagrammatic}) to decompose the terms generated by the monodromy relation (\ref{fundamental_cross_ratio_identity}), resulting in the following, `fundamental' identity,
\eq{1=-\!\sum_{\substack{\alpha\in A^{\phantom{c}}\\\beta\in A^c}}\sum_{\substack{\\[-2pt]\fwboxL{20pt}{\hspace{-0pt}\sigma\!\in\!(A_1\!\shuffle\!{A_2^R})}}}\text{PT}(a,\sigma,\alpha)\z{a}{\alpha}\times(-1)^{|A_2|}\frac{s_{\alpha\beta}}{s_A}\frac{\z{\beta}{b}}{\z{b}{a}\z{\alpha}{\beta}}\,.\label{combined_fundamental_identity_algebraic}}
This relation can be represented diagrammatically as follows:
\eq{\hspace{-90pt}\fwboxR{0pt}{\fig{-27.14pt}{1}{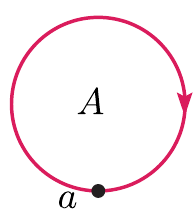}}\;\bigger{=}\;\;\fwboxL{0pt}{\hspace{-5pt}\sum_{\substack{\alpha\in A^{\phantom{c}}\\\beta\in A^c\\[-2.5pt]\fwboxL{20pt}{\sigma\!\!\in\!\!(\!A_1\!\shuffle\!{A_2^R})}}}(-1)^{|A_2|}\frac{s_{\alpha\beta}}{s_A}\fig{-27.14pt}{1}{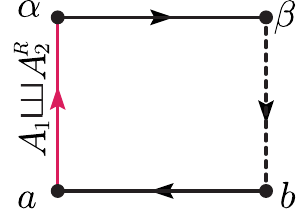}}\label{combined_fundamental_identity_diagrammatic}}

The last ingredient of our algorithm deals with the terms in $\Pfprime\Psi$ involving diagonal entries of the $C$-matrix. We have so-far chosen to represent them as one-cycles abstractly as one-cycles `$\ab{a}$' in the denominator. But in \mbox{ref.\ \cite{Bjerrum-Bohr:2016juj}} it was shown that---up to terms that vanish by gauge invariance and momentum conservation---these entries can always be expanded as follows:
\eq{C_{a\,a}=\sum_{\beta\notin\{a\}}\ek{a}{\beta}\frac{\z{\beta}{b}}{\z{b}{a}\z{a}{\beta}},\label{expansion_of_diagonal_entries_of_c_matrix_algebraic}}
for any choice of $b\!\neq\!a$. Representing this diagrammatically, and using a one-cycle \mbox{$A\!=\!\{a\}$} to represent $C_{a\,a}$ (connecting the point $a$ to itself),
\eq{\hspace{-90pt}\fwboxR{0pt}{\fig{-27.14pt}{1}{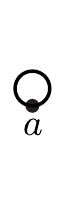}}\;\bigger{=}\;\;\fwboxL{0pt}{\hspace{-5pt}\sum_{\substack{\alpha\in \{a\}{\phantom{c}}\\\beta\in \{a\}^c}}\hspace{-2pt}\ek{\alpha}{\beta}\fig{-27.14pt}{1}{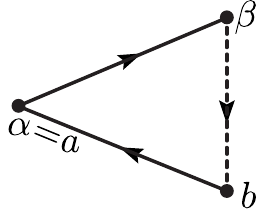}}\label{expansion_of_diagonal_entries_of_c_matrix_diagrammatic}}
its similarity to equation (\ref{combined_fundamental_identity_diagrammatic}) becomes manifest. Indeed, although the term in (\ref{combined_fundamental_identity_algebraic}) involving $\alpha\!=\!a$ vanishes because $\z{a}{a}\!=\!0$, we can view this as the {\it only} term $\alpha\!\in\!A\!\equiv\!\{a\}$ being summed in (\ref{expansion_of_diagonal_entries_of_c_matrix_algebraic}). Thus, we may consider the two relations as being two versions of the same operation: when $|A|\!=\!1$, we use (\ref{expansion_of_diagonal_entries_of_c_matrix_diagrammatic}); and when $|A|\!\geq\!2$, we use (\ref{combined_fundamental_identity_diagrammatic}). To be clear, we may simply write:
\eq{\hspace{-120pt}\fwboxR{0pt}{\fig{-27.14pt}{1}{combined_reduction_0}}\;\bigger{=}\;\;\fwboxL{140pt}{\hspace{-5pt}\sum_{\substack{\alpha\in A^{\phantom{c}}\\\beta\in A^c\\[-2.5pt]\fwboxL{20pt}{\sigma\!\!\in\!\!(\!A_1\!\shuffle\!{A_2^R})}}}(-1)^{|A_2|}\,n_{\alpha,\beta}\fig{-27.14pt}{1}{combined_reduction_1}}\hspace{30pt},\hspace{12pt} n_{\alpha,\beta}\equiv\left\{\begin{array}{l@{$\;\;$}l}s_{\alpha\beta}/{s_A}&|A|>1\\\ek{\alpha}{\beta}&|A|=1\end{array}\right.\!.\hspace{-140pt}\label{fundamental_reduction_equation_diagrammatic}}
Combined in this way, we will refer to this as {\it the fundamental reduction} of the cycle $A$ relative to the points $a\!\in\!A$ and $b\!\notin\!A$. One small subtlety that is worth mentioning here is that equation (\ref{fundamental_reduction_equation_diagrammatic}) still requires that $|A|\!\leq\!(n\mi2)$. Conveniently, whenever there exists a cycle with $|A|\!=\!(n-1)$, we can choose to reduce relative to its complement, for which (\ref{fundamental_reduction_equation_diagrammatic}) applies.

\newpage
\vspace{-0pt}\subsection{The Systematic Reduction of Terms}\label{subsec:algorithm}\vspace{-0pt}

As described above, every term arising in the expansion of the (reduced) Pfaffian $\Pfprime\Psi$ corresponds involves a product of factors $\z{a}{b}$ that can always be considered a product of $\text{PT}$-factors for each Hamiltonian cycle. Let us ignore the kinematic numerators for the moment as they will play no role in our analysis. We would like to show that any such factor can be decomposed, through the iterated use of the {\it fundamental reductions} described above (\ref{fundamental_reduction_equation_diagrammatic}), into sums of terms each involving only a single Hamiltonian cycle. These can then be expanded into any KK-basis using (\ref{kk_relations_algebraic}). As reviewed in \mbox{section \ref{sec:review_of_scattering_equations}}, this will result in a representation that makes manifest the color-kinematics duality of Yang-Mills amplitudes.

Let us now describe the concrete algorithm by which this can be done systematically. The starting point will always be some term appearing in the expansion of the $\Pfprime\Psi$ which involves just a number of disjoint Hamiltonian cycles (including one-cycles). We will refer to any term of this form as being of {\bf Type I}. If there is only a single Hamiltonian cycle of length $n$, no reduction is required. If there is more than one cycle, we simply use the fundamental reduction: choose any cycle $A$ and any two points $a\!\in\!A$ and $b\!\notin\!A$, and reduce the cycle $A$ relative to $a$ and $b$.

Obviously, because $b\!\notin\!A$ and the term involves a union of disjoint cycles, $b$ must lie along some other cycle, $b\!\in\!B$. Thus, the summation over $\beta$ in the fundamental reduction can be organized by whether or not $\beta\!\in\!B$. These two cases can be represented diagrammatically as follows:
\vspace{-5pt}\eq{\hspace{-100pt}\fwboxR{0pt}{\begin{array}{c}\\[-5pt]\fig{-32.14pt}{1}{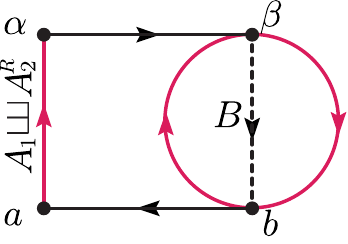}\\[-5pt]\text{{\bf I.a}}\end{array}}\hspace{30pt}\fwboxL{0pt}{\begin{array}{c}\\[-5pt]\fig{-32.14pt}{1}{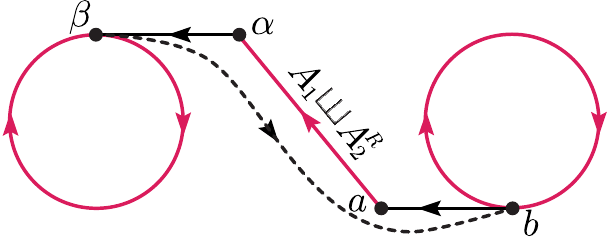}\\[-5pt]\text{{\bf I.b}}\end{array}}\hspace{-50pt}\label{type1_expansions}\vspace{-0pt}}
Here, we have left implicit the kinematic factors and the summations involved in the reduction formula (\ref{fundamental_reduction_equation_diagrammatic}). Also, there can be an arbitrary number of other, disjoint cycles involved---in the figure above, we have only drawn those cycles relevant to particular terms in the expansion. The first of these cases is easily seen to become (a sum of terms) of Type I, upon the use of identity (\ref{kk_relations_diagrammatic}):
\eq{\hspace{-50pt}\fwboxR{0pt}{\fig{-32.14pt}{1}{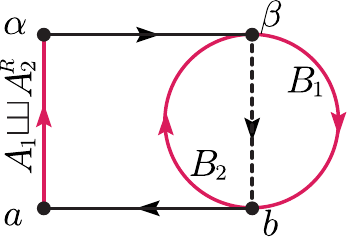}\hspace{10pt}}\bigger{=}\fwboxL{0pt}{\hspace{10pt}\sum_{\substack{\\[-2.5pt]\fwboxL{20pt}{\sigma\!\!\in\!\!(\!B_1\!\shuffle\!{B_2^R})}}}(-1)^{|B_2|}\fig{-27.14pt}{1}{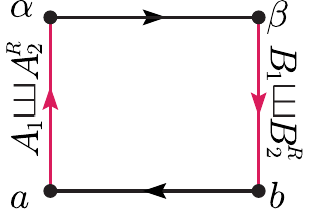}}\label{type1a_reduction}}
Because these terms are individually of Type I, we can reduce them further without encountering any additional complexity relative to that already in (\ref{type1_expansions}).

The second class of terms in (\ref{type1_expansions}) are not of Type I. Indeed, we will refer to integrands such as these as being of {\bf Type II}. More generally, Type II terms are those which include an arbitrary number of disjoint cycles, exactly two of which are connected by a {\it chain} in the following way:
\vspace{-12pt}\eq{\begin{array}{c}\\[-10pt]\fwbox{0pt}{\fig{-27.14pt}{1}{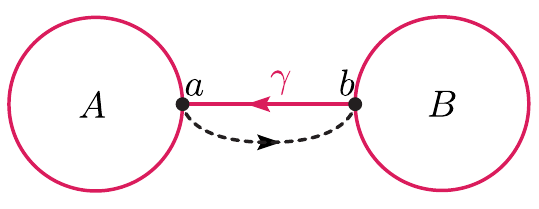}}\\[-10pt]\text{{\bf Type II}}\end{array}\label{definition_of_type_II_terms}}
When the chain $\gamma$ is of length one---when it consists of a single link $\z{b}{a}$---it reduces trivially toType I by canceling the link against the oppositely-oriented numerator. This introduces a minus sign.

Let us now show that for any term of Type II, reduction of $A$ relative to $\{a,b\}$ will lead to (sums of) diagrams either of Type I or Type II. Once this has been shown, it is clear that no further complexity can be generated through iterated reduction; and it will be easy to prove that this procedure will always terminate with terms involving a single Hamiltonian cycle.

In the reduction of the cycle $A$ in (\ref{definition_of_type_II_terms}) relative to $\{a,b\}$, the summation over $\beta$ is naturally organized according to three cases: the first is when $\beta\!\in\!B$; the second, when $\beta$ is along some other cycle; and the third, when $\beta\!\in\!\gamma$. In all three cases, the factor in the denominator $\z{b}{a}$---indicated by a solid line $b\!\to\!a$ in (\ref{fundamental_reduction_equation_diagrammatic})---is cancelled by the factor in the numerator $\z{a}{b}$---indicated by the dashed line in (\ref{definition_of_type_II_terms}); this always introduces an overall minus sign, and effectively replaces the line $b\!\to\!a$ by the chain $\gamma$ connecting $b$ to $a$ via an arbitrary number of other points along the chain. Thus, the three cases can be represented diagrammatically as follows:
\vspace{-16pt}\eq{\hspace{-50pt}\fwbox{500pt}{\fwboxR{0pt}{\begin{array}{c}\\[-5pt]\fig{-32.14pt}{1}{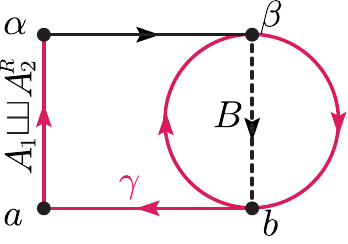}\\\text{{\bf II.a}}\\[-15pt]\end{array}}\fwbox{180pt}{\begin{array}{c}\\[-5pt]\fig{-32.14pt}{1}{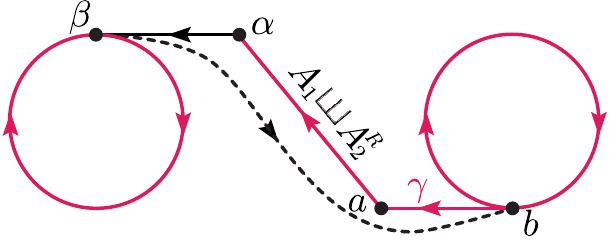}\\\text{{\bf II.b}}\\[-15pt]\end{array}}\fwboxL{0pt}{\hspace{-5pt}\begin{array}{c}\\[-5pt]\fig{-32.14pt}{1}{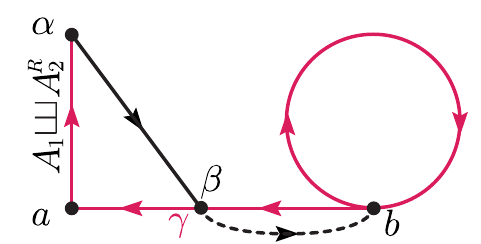}\\\text{{\bf II.c}}\\[-15pt]\end{array}}}\label{type2_expansions}\vspace{5pt}}
The first two of these are structurally identical to the cases resulting from Type I terms given in (\ref{type1_expansions}). And the first of these becomes (a sum of terms of) Type I in exactly the same way as in (\ref{type1a_reduction}):
\vspace{5pt}\eq{\hspace{-50pt}\fwboxR{0pt}{\fig{-32.14pt}{1}{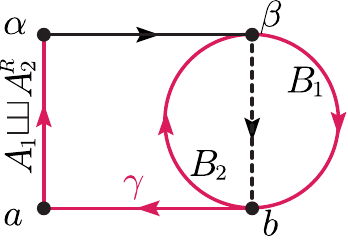}\hspace{10pt}}\bigger{=}\fwboxL{0pt}{\hspace{10pt}\sum_{\substack{\\[-2.5pt]\fwboxL{20pt}{\sigma\!\!\in\!\!(\!B_1\!\shuffle\!{B_2^R})}}}(-1)^{|B_2|}\fig{-27.14pt}{1}{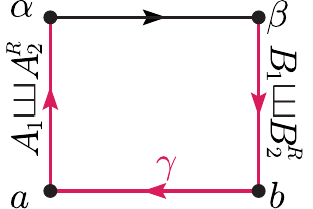}}\label{type2a_reduction}}

The only novelty that appears upon reducing a term of Type II, therefore, is the case II.c shown in (\ref{type2_expansions}). But this is manifestly of Type II---only now, involving a chain $\gamma'$ (the part of $\gamma$ from $b$ to $\beta$) of shorter length. As before, if this new chain consists of a single link, it cancels against the numerator (introducing a minus sign), resulting in a term of Type I.

\vspace{-0pt}\subsection{Proof of Recursive Reduction to Single Cycles}\label{subsec:algorithm_proof}\vspace{-6pt}

From the discussion above, it is not hard to see that we can systematically reduce all terms appearing in the expansion of $\Pfprime\Psi$ into those involving single cycles. And these can be expanded into any KK-basis using (\ref{kk_relations_algebraic}) to obtain a representation of $\Pfprime\Psi$ of the form required (\ref{kk_expansion_of_pfaffian}). Although nearly trivial, let us quickly prove that this is the case.

As shown above, any term of Type I or II can be reduced into sums of terms of the same types. For both of the cases resulting from the reduction of terms of Type I, (\ref{type1_expansions}), the result will be a configuration with one fewer cycle---the cycle $A$ either merges with that of $B$ (I.a), or it becomes part of a chain connecting $B$ to some other cycle (I.b). And the same is true for the first two cases resulting from the reduction of terms of Type II. Only case II.c in (\ref{type2_expansions}) does not lower the total number of cycles; however, these terms are guaranteed to involve terms of Type II connected by {\it shorter} chains $\gamma'$.

Therefore, if we characterize any terms of these two types by the number of cycles, $n_c$, and the length, $|\gamma|$, of any chain between cycles (when no chain exists, $|\gamma|\!=\!0$), then the reduction procedure described above is guaranteed to result in terms for which the pair $\{n_c,|\gamma|\}$ is decreased (lexicographically). Because the sequence of $\{n_c,|\gamma|\}$ characteristics for terms generated by the reductions described above are monotonically decreasing, it is clear that they must terminate in terms with $n_c\!=\!1$ and $|\gamma|\!=\!0$, which cannot be further reduced. \hspace{\fill}Q.E.F.

\newpage
\vspace{-0pt}\section{{Application}: the Four-Particle Amplitude in Yang-Mills}\label{sec:examples}\vspace{-6pt}

The algorithm above clearly reduces all terms appearing the the expansion of $\Pfprime\Psi$ into the form of (\ref{kk_expansion_of_pfaffian}) required to make color-kinematics duality of amplitudes manifest. While sufficiently easy to implement on a computer, for example, it is worthwhile to illustrate how it works in practice for at least one concrete example. Doing so will allow us to highlight several important features of the algorithm. Most notably, it will allow us to highlight the flexibility available at each successive stage of reduction, and the implications of the existence of so many reduction pathways.

The simplest non-trivial case is that of four particles. Direct expansion of $\Pfprime\Psi$, eliminating the columns $\{i,j\}\!=\!\{1,4\}$, results in the following terms:
\eq{\Pfprime\Psi=\frac{n_1}{\ab{1234}}+\frac{n_2}{\ab{1324}}+\frac{n_3}{\ab{14}\ab{23}}+\frac{n_4}{\ab{124}\ab{3}}+\frac{n_5}{\ab{134}\ab{2}}+\frac{n_6}{\ab{14}\ab{2}\ab{3}}\,.\label{four_particle_pf_terms}}
(Recall the notation defined in (\ref{parke_taylor_factor_defined}) above.) Here, the initial 15 terms in the expansion have been organized into groups according to their $z$-dependence. These follow immediately from the definition of $\Pfprime\Psi$ in (\ref{definition_of_reduced_pfaffian_and_of_psi}); for the sake of reference, they are:
\eq{\hspace{-80pt}\begin{array}{lll@{$\;\;\;\;\;\;\;\;\;$}lll}n_{1}&\equiv&\e{34}(s_{23}\,\e{12}\mi \ek{1}{2}\,\ek{2}{3})\pl\ek{4}{3}(\e{23}\,\ek{1}{2}\mi \e{12}\,\ek{3}{2});&n_{4}&\equiv&\e{12}\,\ek{4}{2}\mi \e{24}\,\ek{1}{2};\\n_{2}&\equiv&\e{24}(s_{23}\,\e{13}\mi \ek{1}{3}\,\ek{3}{2})\pl\ek{4}{2}(\e{23}\,\ek{1}{3}\mi \e{13}\,\ek{2}{3});&n_{5}&\equiv&\e{13}\,\ek{4}{3}\mi \e{34}\,\ek{1}{3};\\n_{3}&\equiv&\e{14}(\ek{2}{3}\,\ek{3}{2}\mi s_{23}\,\e{23});&n_{6}&\equiv&\mi\e{14}\hspace{56.5pt}.\end{array}\hspace{-50pt}\label{definition_of_initial_4pt_numerators}}
It is worth mentioning that these numerators can be nicely organized into the $W$- and $U$-notation of \mbox{ref.\ \cite{Lam:2016tlk}}; but their explicit form will play no role in our analysis.

Clearly, the first two terms in (\ref{four_particle_pf_terms}) are already reduced, and already in the desired KK-basis. Therefore, we need only reduce the terms proportional to $n_{i}$ for $i\!=\!3,\ldots,6$. Let us consider each in turn.

Let us start with the first term that requires reduction---that proportional to $n_3$ in (\ref{four_particle_pf_terms}). This term involves two 2-cycles, $\ab{14}$ and $\ab{23}$, and so already we are faced with choices---both of which cycle to reduce, and the points $\{a,b\}$ with which to do the reduction. Of all the possible choices involved, only two result in (essentially) different expressions. Choosing to reduce the cycle $\ab{14}$ with respect to the points $\{a,b\}\!\equiv\!\{1,3\}$, we find the reduction:
\eq{\frac{1}{\ab{14}\ab{23}}=-\frac{s_{24}}{s_{14}}\frac{1}{\ab{1324}}\,;\label{term3_reduction_1}}
choosing instead to reduce the cycle $\ab{23}$ with respect to the points $\{a,b\}\!\equiv\!\{3,4\}$ we would find
\eq{\frac{1}{\ab{14}\ab{23}}=-\frac{s_{12}}{s_{23}}\frac{1}{\ab{1234}}\,.\label{term3_reduction_2}}
The equality between the expressions (\ref{term3_reduction_1}) and (\ref{term3_reduction_1}) is literally a manifestation of the BCJ relations of the amplitude \cite{BCJ}. And this example illustrates an important and general feature of the reduction algorithm: not only are there many choices for how to reduce a term, but these choices multiply at every stage of recursion, resulting in myriad possibilities for the ultimate expressions obtained.

The next two terms, those proportional to $n_{4}$ and $n_{5}$ in (\ref{four_particle_pf_terms}), have no such choices for their reduction: because we must always (and without loss of generality) choose the cycle being reduced to have fewer than $(n\mi1)$ elements, we must reduce the one-cycles of each. (The choice $b$ does not affect either of the reductions in this case.) Therefore, reducing the cycle $\ab{3}$ of the first and $\ab{2}$ of the second, we obtain the following:
\eq{\begin{array}{lllll}\displaystyle\frac{1}{\ab{124}\ab{3}}&=&\displaystyle\frac{\ek{3}{1}}{\ab{1243}}-\frac{\ek{3}{2}}{\ab{1234}}&=&\displaystyle-\left(\frac{\ek{3}{1}+\ek{3}{2}}{\ab{1234}}+\frac{\ek{3}{1}}{\ab{1324}}\right);\\[12pt]
\displaystyle\frac{1}{\ab{134}\ab{2}}&=&\displaystyle\frac{\ek{2}{1}}{\ab{1243}}-\frac{\ek{2}{3}}{\ab{1324}}&=&\displaystyle-\left(\frac{\ek{2}{1}}{\ab{1234}}+\frac{\ek{2}{1}+\ek{2}{3}}{\ab{1324}}\right).\end{array}\label{tems45_reductions}}
Here, we have used the KK-relations, (\ref{kk_relations_algebraic}), to expand $\text{PT}(1,2,4,3)$ as,
\eq{\frac{1}{\ab{1243}}=-\left(\frac{1}{\ab{1234}}+\frac{1}{\ab{1324}}\right)\,,\label{explicit_kk_reduction_of_1243}}
for each of the reduced expressions in (\ref{tems45_reductions}).

The final term in (\ref{four_particle_pf_terms}), proportional to $n_6$, has the most possible variation for its reduction. Many of these choices are essentially the same. For example, we can start by reducing the cycle $\ab{3}$ with respect to the points $\{a,b\}\!\equiv\!\{3,1\}$; the result is:
\eq{\frac{1}{\ab{14}\ab{2}\ab{3}}=-\frac{\ek{3}{1}}{\ab{134}\ab{2}}-\frac{\ek{3}{2}}{\ab{14}\ab{2}}\frac{\z{2}{4}}{\z{4}{3}\z{3}{2}}\,.\label{reduction_of_term_6_step_1}}
Notice that the second term above is of Type II according to the classification in our algorithm; the additional factor represents an (oriented) chain from $4\!\to\!3\!\to\!2$. The first term in (\ref{reduction_of_term_6_step_1}) reduces uniquely, resulting in the expression given in (\ref{tems45_reductions}); the second term is more interesting. Because it is of Type II, there are no choices for its reduction and we are forced to reduce the cycle $\ab{14}$ with respect to the points $\{a,b\}\!\equiv\!\{4,2\}$. This results in,
\eq{\frac{1}{\ab{14}\ab{2}}\frac{\z{2}{4}}{\z{4}{3}\z{3}{2}}=-\frac{\ek{2}{1}}{\ab{1234}}-\frac{\ek{2}{3}}{\ab{14}\ab{23}}\,.\label{reduction_of_term_6_2_step_2}}
Notice that the second term is exactly that already encountered above---with two choices for its ultimate reduction, (\ref{term3_reduction_1}) and (\ref{term3_reduction_2}). Using the first of these possible reductions for this last term, and combining everything together, we find that
\eq{\frac{1}{\ab{14}\ab{2}\ab{3}}=\frac{\ek{2}{1}(\ek{3}{1}\!\pl\ek{3}{2})}{\ab{1234}}\pl\left(\ek{3}{1}(\ek{2}{1}\!\pl\ek{2}{3})-\frac{s_{24}\,\ek{2}{3}\,\ek{3}{2}}{s_{14}}\right)\frac{1}{\ab{1324}}\,.\label{term6_full_reduction_v1}}
If we had instead used (\ref{term3_reduction_2}) to expand the last term in (\ref{reduction_of_term_6_2_step_2}), we would have had:
\eq{\frac{1}{\ab{14}\ab{2}\ab{3}}=\left(\ek{2}{1}(\ek{3}{1}\!\pl\ek{3}{2})-\frac{s_{12}\,\ek{2}{3}\,\ek{3}{2}}{s_{14}}\right)\frac{1}{\ab{1234}}\pl\frac{\ek{3}{1}(\ek{2}{1}\!\pl\ek{2}{3})}{\ab{1324}}\,.\label{term6_full_reduction_v2}}

Combining everything from the work above, and using (\ref{term3_reduction_1}) and (\ref{term6_full_reduction_v1}) for the expansions of the third and sixth terms, respectively, we arrive at the manifestly color-kinematics dual representation for the Pfaffian:
\eq{\Pfprime\Psi ~\equiv~ \frac{n_{1,\{2,3\},4}}{\ab{1234}}+\frac{n_{1,\{3,2\},4}}{\ab{1324}}\,\,,\label{manifestly_dual_form_of_4pt_pf}}
where
\begin{align}\hspace{-10pt}n_{1,\{2,3\},4}&\equiv n_1\mi n_4\,(\ek{3}{1}\!\pl\ek{3}{2})\mi n_5\,\ek{2}{1}\!\pl n_6\,\ek{2}{1}(\ek{3}{1}\!\pl\ek{3}{2});\label{kk_dual_numerators_for_4pt}\\
\hspace{-10pt}n_{1,\{3,2\},4}&\equiv n_2\mi n_3\frac{s_{24}}{s_{14}}\mi n_4\,\ek{3}{1}\mi n_5(\ek{2}{1}\!\pl\ek{2}{3})\pl n_6\Big(\ek{3}{1}(\ek{2}{1}\!\pl\ek{2}{3})\mi\frac{s_{24}\,\ek{2}{3}\,\ek{3}{2}}{s_{14}}\Big)\,.\nonumber
\end{align}

Let us conclude with some observations about the above example. The form of the color-kinematic-dual numerators (\ref{kk_dual_numerators_for_4pt}) happens to be in a so-called `local' form---with no denominators of the numerators already appearing in the corresponding term's denominators (although this fact is not manifestly obvious as written). Also, if instead of using the reduction (\ref{term3_reduction_1}) for both appearances of this term (for $n_3$ and $n_6$) we had taken the average of the expressions (\ref{term3_reduction_1}) and (\ref{term3_reduction_2}), we would have found that $n_{1,\{2,3\},4}$ and $n_{1,\{3,2\},4}$ would have been manifestly permutations of each other \cite{Broedel:2011pd}---meaning, permuting the labels $\{2,3\}\!\leftrightarrow\!\{3,2\}$ would have exchanged the two expressions. We do not know how general these facts are beyond $n\!=\!4$, but it seems definitely worthwhile to explore the space of color-kinematic-dual formulae that result from different pathways through our reduction algorithm.

\vspace{-6pt}\section{Conclusions}\label{sec:conclusions}\vspace{-6pt}

Using the integration rules of \mbox{refs.\ \cite{Baadsgaard:2015voa,Baadsgaard:2015ifa,Baadsgaard:2015hia}} augmented by those of \cite{Bjerrum-Bohr:2016juj}, the CHY-formalism provides a neat and compact expression for $n$-point Yang-Mills amplitudes in any number of dimensions. However, it is not directly given in a form that makes color-kinematics duality manifest. In this paper, we have shown that a systematic reduction algorithm puts all CHY-integrands in a form corresponding to precisely two Parke-Taylor factors, one shuffled with respect to the other. This trivializes all integrations and puts the amplitude manifestly in a form corresponding to trivalent Feynman graphs, dressed with non-trivial numerators that depend on the polarizations and momenta. Once in this form, the amplitude can be further reduced down to a KK-basis by means of the algebraic part of the monodromy relations. As follows from the analysis of \mbox{ref.\ \cite{Cachazo:2013iea}}, those numerators satisfy Jacobi relations and therefore imply color-kinematics duality of the scattering amplitude.

Since our algorithm provides a straightforward and explicit way to construct BCJ-numerators for Yang-Mills theory amplitudes with any number of external legs, we expect that it may shed new light on some of the issues discussed in \cite{Tye:2010dd,BjerrumBohr:2010zs,Feng:2010my,Chen:2011jxa,Broedel:2011pd,Bern:2011ia,Mafra:2011kj,Du:2013sha,Mafra:2015vca}. It is also interesting to note that by numerator squaring we immediately recover $n$-point gravity amplitudes \cite{Bern:2010ue,Bern:2010yg}.

Because our reduction proof is entirely independent of the detailed factors that  dress the auxiliary integration variables of the CHY-formalism, it applies to many other cases. Most importantly, with the help of the algorithm described here one can merge Hamiltonian cycles into larger cycles. To illustrate this, one can consider the interesting identities recently derived by Stieberger and Taylor \cite{Stieberger:2016lng} that involve combinations of Einstein-Yang-Mills theory and pure Yang-Mills theory (see also \mbox{\cite{Chen:2010ct}}). These relations, and generalizations thereof, have been shown to follow from the scattering equation formalism \cite{Nandan:2016pya,delaCruz:2016gnm}. The identities needed to show this are precisely special cases of the general reduction algorithm derived in this paper. Indeed, it is clear that these relations are nothing but special cases of our general algorithm. Applying our algorithm to the most general mixed gluon-graviton amplitudes, we can rewrite those as a sum over pure Yang-Mills theory amplitudes. In fact, our algorithm indicates that amplitudes of any theory with a factor $\Pfprime\Psi$ in its CHY-integrand can be written as a linear combination of Yang-Mills amplitudes---for example, Born-Infeld theory. Going one level deeper, we note that the basic building blocks are invariable the trivalent graphs of $\phi^3$-theories, as described in \mbox{ref.\ \cite{Baadsgaard:2015ifa}}. Only different `dressing factors' on those cubic vertices distinguish these classes of theories. Remarkably, this holds for pure gravity as well. It would be very interesting to understand the interpretation of these relations, manifest in the scattering equation formalism, directly within traditional quantum field theory and string theory.

\section*{Acknowledgements}\vspace{-6pt}

The authors would like to thank NORDITA for their hospitality during the program ``Aspects of Amplitudes'', and the Isaac Newton Institute for Mathematical Sciences during the program ``Gravity, Twistors, and Amplitudes''---which was supported by EPSRC grant number EP/K032208/1. This work was supported in part by the Danish National Research Foundation (DNRF91), by a MOBILEX research grant from the Danish Council for Independent Research (JLB), and by Qiu-Shi and Chinese NSF funding under contract Nos.\ 11575156, 11135006, and 11125523 (BF).

\newpage
\providecommand{\href}[2]{#2}\begingroup\raggedright\endgroup

\end{document}